\documentclass[12pt]{article}

\usepackage{amsmath,amssymb}

\oddsidemargin  - 5mm
\evensidemargin - 5mm

\newcommand{\be}{\begin{equation}}
\newcommand{\ee}{\end{equation}}
\newcommand{\bea}{\begin{eqnarray}}
\newcommand{\eea}{\end{eqnarray}}

\newcommand{\Gr}{Gr\"obner\,}

\begin{document}

\title{\bf
Analysis of constraints in light-cone version of $SU(2)$ Yang-Mills mechanics}

\author{
V. P. Gerdt  $^a$,
A. M. Khvedelidze $^{b,\ c}$ and
D. M. Mladenov $^c$\\
$^a$ {\it Laboratory of Information Technologies},\\
Joint Institute for Nuclear Research, 141980 Dubna, Russia\\
$^b$ {\it A. Razmadze Mathematical Institute},\\
GE-380093, Tbilisi, Georgia \\
$^c$ {\it Bogoliubov Laboratory of Theoretical Physics},\\
Joint Institute for Nuclear Research, 141980 Dubna, Russia}

\date{\empty}
\maketitle

\begin{abstract}
We study the classical dynamics of mechanical model obtained from
the light-cone version of $SU(2)$ Yang-Mills field theory
under the supposition of gauge potential dependence only on ``time''
along the light-cone direction.
The computer algebra system Maple was used strongly to compute and separate the
complete set of constraints.
In contrast to the instant form of Yang-Mills mechanics
the constraints here represent a mixed form of first and
second-class constraints and reduce the number of the physical
degrees of freedom up to four canonical one.
\end{abstract}


\section{Introduction}


Notion of the evolution of observables is the key element
in analyzing of the physical properties of any relativistic field theory.
After Dirac's famous work entitled
{\it ``Forms of Relativistic Dynamics''} \cite{Dirac1949}
it has been recognized that the different choices of the
time evolution parameter can drastically change the
content and interpretation of the theory.
The simplest and well-known example illustrated this
observation is the light-cone dynamics of free scalar field.
In contrast to the corresponding instant time model,
in this case, owing to the choice of time evolution parameter along the
light-cone characteristics, theory becomes degenerate, the
corresponding Hessian is zero \cite{Sundermeyer}.
Dealing with gauge theories on the light-cone
we encounter much more complicated description
than for the ordinary instant form dynamics
(see e.g. recent reviews \cite{Heinzl:2001ht}-\cite{Srivastava:1999js}).

In the present talk we would like to state some results
concerning the light-cone description of
simple mechanical model originated from the $SU(2)$ Yang-Mills theory
under assumption of spatial homogeneity of the fields on the light-cone.
This means that we shall consider the light-cone action for $SU(2)$
Yang-Mills model with the gauge potential only light-cone time depending.
The dynamical system, obtained under such a supposition
contain finite number of degrees of freedom and possesses gauge invariance.
Our aim is to study its Generalized Hamiltonian dynamics
\cite{Sundermeyer,DiracL,HenTeit}
and to compare it with the corresponding instant form of the Yang-Mills mechanics,
intensively studied during the last decades
(see e.g \cite{MatSav}-\cite{KM} and references therein).

Using the Generalized Hamiltonian formalism
for degenerate systems \cite{Sundermeyer,DiracL,HenTeit} and
exploiting the Maple package~\cite{GG99} implementing
algorithm Dirac-\Gr for computation and separation of constraints
we found the complete set of constraints and performed their separation
into sets of first and second-class constraints.

Our calculations demonstrate that the light-cone version
of Yang-Mills mechanics differs from its instant form
counterpart in the character of the local gauge invariance
and therefore the corresponding unconstrained Hamiltonian
systems describe different canonically non-equivalent models.


\section{Description of the model}


Let us start with a general formulation of the Yang-Mills theory
on four-dimensional Minkowski space $M_4$,
endowed with some metric $g$, tensor field of type $(0,2)$,
\begin{equation} \label{eq:metric}
g = g_{\mu\nu}\, \omega^\mu\,  \otimes \, \omega^\nu \,.
\end{equation}
At every point point $P \in M_4$ we use a basis $\{\omega^\mu \,, \mu = 0,1,2,3\}$
of 1-forms in the cotangent space $T_P^\ast(M_4)$.
The metric $g$ defines an inner product between two vectors in the tangent space
$T_P(M_4)$ and if one fixes a basis $e_\mu$ in $T_P(M_4)$,
dual to the basis of 1-forms $\omega^\mu$, the components of the metric tensor are given as
\begin{equation}
g_{\mu\nu} = g( e_\mu, e_\nu )\,.
\end{equation}

Using these geometrical settings, the action of the Yang-Mills field theory
can be represented in coordinate free form
\begin{equation}
\label{eq:gaction}
I : = \frac{1}{g^2}\, \int_{M_4} \mbox{tr} \,  F\wedge * F \,.
\end{equation}
Here the $SU(2)$ algebra valued curvature 2-form
\begin{equation}
F = d A +  A \wedge A
\end{equation}
is constructed from the connection 1-form $A = A_\mu \, \omega^\mu$.
The connection and the curvature as Lie algebra valued
quantities are expressed in terms of the Pauli matrices
$\sigma^a \,, a = 1,2,3$~
\footnote{
The Pauli matrices satisfy
$$
[\sigma^a, \sigma^b] = 2\, i \, \varepsilon^{abc}\, \sigma^c \,,
\quad \quad
\mbox{tr}\, \sigma^a \sigma^b = 2 \,\delta^{ab}\,.
$$
}
\begin{equation}
A = A^a \, \frac{\sigma^a}{2 i}\,, \qquad F = F^a \, \frac{\sigma^a}{2 i}\,\,.
\end{equation}
The metric $g$ enters the action through the dual field strength
tensor defined in accordance to the Hodge star operation
\begin{equation}
* F_{\mu\nu}  = \frac{1}{2}\, \epsilon_{\mu\nu\alpha\beta}\, F^{\alpha\beta}\,.
\end{equation}

If one fixes Lorentzian coordinates in Minkowski space $M_4$
$x^\mu = (x^0, x^1, x^2, x^3)$ and choose a coordinate basis for the tangent vectors
$e_\mu = \frac{\partial}{\partial x^\mu}$,
we have the conventional Minkowskian metric
$\eta = \|1, -1, -1, -1 \|$ and
the corresponding action (\ref{eq:gaction}) will provide Yang-Mills equations in
the instant form with a time variable $t = x^0$.

To formulate the light-cone version of the theory let us
introduce basis vectors in the tangent space $T_P(M_4)$
\begin{equation}
e_{\pm} := \frac{1}{\sqrt 2} \, \left( e_0 \pm e_3 \right) \,, \quad
e_\bot :=  \left( e_k \,, k = 1, 2 \right) \,.
\end{equation}
The first two vectors are tangent to the light-cone
and the corresponding coordinates are referred usually as the light-cone coordinates
$
x^\mu = \left(x^+, x^-, x^\bot\right)
$
with
\begin{equation}
x^\pm := \frac{1}{\sqrt 2}\, \left( x^0 \pm x^3 \right) \,, \qquad
x^\bot :=  \left(x^k \,, k = 1, 2 \right) \,.
\end{equation}
The light-cone basis vectors $\left(e_\pm, e_k \right)$
\footnote{
Hereinafter the latin indexes $i,j,k$ run over 1,2.
Further we shall treat in equal footing the up and down isotopic indexes,
denoted with $a, b, c, d$.}
determine, according to (\ref{eq:metric}),
the so-called light-cone metric, whose non-zero elements are
\begin{equation}
g_{+-} = g_{-+} = - g_{11} = - g_{22} = 1
\end{equation}
and thus the connection 1-form  in the light-cone formulation is given as
\begin{equation} \label{eq:conlc}
A = A_+ \, dx^+ + A_- \, dx^- + A_k \, dx^k \,.
\end{equation}

Now we are ready to define the Lagrangian corresponding to the light-cone Yang-Mills mechanics.
By definition the Lagrangian of Yang-Mills mechanics
follows from the Lagrangian of Yang-Mills theory if one suppose that
connection 1-form $A$ depends only on light-cone ``time variable'' $x^+$
\begin{equation}
A = A(x^+) \,.
\end{equation}
Using the definition (\ref{eq:gaction})
and (\ref{eq:conlc}) we find the Lagrangian of the Yang-Mills light-cone mechanics
\begin{equation} \label{eq:lagr}
L : = \frac{1}{2 g^2} \,
\left(
F^a_{+ -} \,  F^a_{+ -} + 2 \, F^a_{+ k} \, F^a_{- k} - F^a_{12} \, F^a_{12}
\right)\,,
\end{equation}
where the field-strength tensor light-cone components are
\begin{equation}
\left\{
\begin{array}{l}
F^a_{+ -} = \frac{\partial A^a_-}{\partial x^+} + \epsilon^{abc}\, A^b_+ \,  A^c_- \,,\\[0.2cm]
F^a_{+ k} = \frac{\partial A^a_k}{\partial x^+} + \epsilon^{abc}\, A^b_+ \,  A^c_k \,, \\[0.2cm]
F^a_{- k} = \epsilon^{abc} \,  A^b_- \, A^c_k \,, \\[0.2cm]
F^a_{i j} = \epsilon^{abc}\, A^b_i \, A^c_j\,.
\end{array}
\right.    \label{eq:components}
\end{equation}


\section{
Hamiltonian formulation of $SU(2)$ Yang-Mills \\ mechanics on the light-cone}
\label{sec:YMLC1}


In this section we present the main results of this paper.
The underlying computations were done with the Maple package implementing algorithm
Dirac-\Gr \cite{GG99} for computation and separation of constraints for
Lagrangian dynamical systems of polynomial type.
Some computational details are described in the next section.

The choice of the light-cone time variable
\begin{equation}
\tau = x^+
\end{equation}
as the evolution parameter prescribes a Legendre transformation of the dynamical variables
$\left(A_+, A_-,  A_k \right)$
\footnote{
To simplify the formulas we shall use overdot to denote
derivative of functions with respect to light-cone time variable $x^+$.}
\begin{equation} \label{eq:legendre}
\left\{
\begin{array}{l}
\pi^-_a  =  \frac{\partial L}{\partial \dot{A^a_-}} =
\frac{1}{g^2} \, \left( \dot{A^a_- } + \epsilon^{abc} \, A^b_+ \, A^c_- \right) \,, \\[0.2cm]
\pi_a^k = \frac{\partial L}{\partial \dot{A^a_k}} =
\frac{1}{g^2} \, \epsilon^{abc} \, A^b_- \, A^c_k \,.
\end{array}
\right.
\end{equation}

\noindent
For this set of equations the designed in~\cite{GG99} Dirac-\Gr algorithm leads
to the primary constraints
\begin{eqnarray}
&& \varphi^{(1)}_a   := \pi^+_a = 0 \,,\label{eq:prcon-1}\\
&& \chi^a_k := g^2 \, \pi^a_k  + \epsilon^{abc} \, A^b_- A^c_k=0\,. \label{eq:prcon-2}
\end{eqnarray}
Then, the canonical Hamiltonian is given by
\begin{equation} \label{eq:khrlh}
H_C = \frac{g^2}{2}\,  \pi^-_a  \pi^-_a - \,
\epsilon^{abc} \,  A^b_+ \left(A^c_- \, \pi^-_a  + A^c_k \,\pi^k_a \right) +
V(A_k)
\end{equation}
with a potential term in (\ref{eq:khrlh})
\begin{equation}
V(A_k) = \frac{1}{2 g^2} \,
\left[
\left(A^b_1 A^b_1\right)\, \left(A^c_2 A^c_2 \right) -
\left(A^b_1 A^b_2\right)\, \left(A^c_1 A^c_2 \right)
\right] \,.
\end{equation}
The nonvanishing Poisson brackets are
\begin{eqnarray}
&& \{ A^a_\pm \,, \pi^\pm_b \} = \delta^a_b \,,\\
&& \{ A^a_k \,,  \pi_b^l \}= \delta_k^l \delta^a_b \,.
\end{eqnarray}
With respect to these fundamental Poisson brackets the primary constraints
$\left( \varphi^{(1)}_a,\chi^a_k\right)$ obey the algebra
\begin{eqnarray}
&& \{ \varphi^{(1)}_a \,, \varphi^{(1)}_b\} = 0 \,,\\
&& \{ \varphi^{(1)}_a \,, \chi^b_k \} = 0 \,,\\
&& \{ \chi^a_i \,, \chi^b_j\} = -2\, g^2 \epsilon^{abc}\, A^c_- \, g_{i j} \,.
\end{eqnarray}
According to Dirac's prescription, the dynamics for degenerate
theories is governed by total Hamiltonian, which differs from the canonical
one by linear combination of the primary constraints.
In case of the light-cone Yang-Mills mechanics the total Hamiltonian
has the form
\begin{equation} \label{eq:toth}
H_T = H_C - 2 \, \mbox{tr} \left(U (\tau)\, \varphi^{(1)} \right) -
2 \, \mbox{tr}\left(V_k(\tau)\, \chi_k \right)\,,
\end{equation}
where $U(\tau)$ and $V_k(\tau)$ are arbitrary $SU(2)$ valued functions of
the light-cone time $\tau = x^+$.
Using this Hamiltonian it is necessary to check the
dynamical self-consistence of the primary constraints.
The requirement of conservation in time of the primary
constraints $\varphi^{(1)}_a$ leads to the equations
\begin{equation}
0 = \dot \varphi^{(1)}_a = \{\pi^+_a\,, H_T\} =
\epsilon^{abc}\, \left(A^b_-  \pi^-_c  +  A^b_k \pi^k_c \right)\,.
\end{equation}
Therefore there are three secondary constraints $\varphi^{(2)}_a$
\begin{equation} \label{eq:secgauss}
\varphi^{(2)}_a := \epsilon_{abc}
\left(A^b_-  \pi^-_c  +  A^b_k \pi^k_c \right)=0\,,
\end{equation}
which obey the $SO(3, \mathbb{R})$ algebra
\begin{equation}
\{ \varphi^{(2)}_a \,, \varphi^{(2)}_b \} = \epsilon_{abc}\, \varphi^{(2)}_c \,.
\end{equation}
The same procedure for the primary constraints
$\chi^a_k$ gives
\begin{equation} \label{eq:secchi}
0 = {\dot\chi}^a_k =
\{\chi^a_k\,, H_C \} - 2\, g^2\, \epsilon^{abc} \, V^b_k\, A^c_-   \,.
\end{equation}
Because the matrix $ \| \epsilon^{abc}\, A^c_- \| $
is degenerate, its rank is
\begin{equation}
\mbox{rank} \|\, \epsilon^{abc}\, A^c_- \,\| = 2\,,
\end{equation}
one can determine among the Lagrange multipliers $V_b^k$ only four ones.
Using the unit vector
\begin{equation}
n^a = \frac{A^a_-}{ \sqrt{(A_-^1)^2 + (A_-^2)^2 + (A_-^3)^2} }\,,
\end{equation}
which is the null vector of the matrix $\| \varepsilon^{abc} \, A^c_- \,\|$,
one can decompose the six primary constraints $\chi^a_k$
\begin{eqnarray}
&& \chi^a_{k \bot} :=
\chi^a_k - \left( n^b \chi^b_k  \right) \, n^a \,, \\
&& \psi_k : = n^a \chi^a_k  \,.
\end{eqnarray}
Constraints $\chi^a_{k \bot}$ are functionally dependent due to the conditions
\begin{equation} \label{eq:dependence}
n^a \, \chi^a_{k \bot} = 0
\end{equation}
and choosing among them any four independent constraints
we are able to determine four Lagrange multipliers $V^k_{\ b \bot}$.
The two constraints $\psi_k$ satisfy the Abelian algebra
\begin{equation}
\{ \psi_i \,, \psi_j \} = 0 \,.
\end{equation}
One can verify that the Poisson brackets
of $\psi_k$ and $\varphi^{(2)}_a$ with the total Hamiltonian on the constraint surface (CS)
are zero
\begin{eqnarray} \label{eq:check}
&& \{ \psi_k \,, H_T \}_{\,\vert{CS}} = 0 \,, \\
&& \{ \varphi^{(2)}_a \,, H_T \}_{\,\vert{CS}} = 0
\end{eqnarray}
and thus there are no ternary constraints.
To summarize: we arrive at the set of constraints
$\left(\varphi^{(1)}_a, \psi_k, \varphi^{(2)}_a, \chi^b_{k \bot}\right)$
with Poisson bracket relations between the constraints
$\left(\varphi^{(1)}_a, \psi_k, \varphi^{(2)}_a\right)$
\begin{eqnarray}
&& \{ \varphi^{(1)}_a \,, \varphi^{(1)}_a\} = 0 \,, \label{eq:group_1} \\
&& \{ \psi_i \,, \psi_j \} = 0 \,, \label{eq:group_2} \\
&& \{ \varphi^{(2)}_a \,, \varphi^{(2)}_b\} = \epsilon_{abc}\, \varphi^{(2)}_c \,, \label{eq:group_3} \\
&& \{ \varphi^{(1)}_a \,, \psi_k\} =  \{\varphi^{(1)}_a \,,\varphi^{(2)}_b \} =
\{ \psi_k \,, \varphi^{(2)}_a \} = 0 \,. \label{eq:group_4}
\end{eqnarray}
The remaining constraints $\chi^b_{k \bot}$ obey the relations
\begin{equation} \label{eq:bracket-1}
\{ \chi^a_{i \bot} \,, \chi^b_{j \bot} \} =
- 2 \, g^2 \, \epsilon^{abc} \, A^c_- \, g_{i j}
\end{equation}
and the Poisson brackets between these two sets of constraints are
\begin{eqnarray}
&&
\{\varphi^{(2)}_a \,, \chi^b_{k \bot} \} =
\epsilon^{abc} \, \chi^c_{k \bot} \,, \label{eq:bracket-2}\\
&& \{ \varphi^{(1)}_a \,, \chi^b_{k \bot}\} =
\{ \psi_i \,, \chi^b_{j\bot}\} = 0 \,. \label{eq:bracket-3}
\end{eqnarray}
From this algebra of constraints we conclude that we have eight first-class constraints
$\left(\varphi^{(1)}_a, \psi_k, \varphi^{(2)}_a \right)$
and four second-class constraints $\chi^a_{k \bot}$.
According to counting of the degrees of freedom eliminated by all these constraints,
after reduction to the unconstrained phase space,
instead of $24$ degrees of freedom possessing the Yang-Mills mechanics on the
light-cone we arrive at
$24 - 4 - 2 (3 + 3 + 2) = 4$ unconstrained degrees of freedom.

Thus one can conclude that in contrast to the instant form of the Yang-Mills mechanics,
where the number of the unconstrained canonical pairs is $12$,
in the light-cone version we have only $4$ physical canonical variables.
It is important to note that such a decreasing of the numbers of the physical
coordinates has two reasons:
as well as the presence of the second-class constraints as the additional
first-class constraints.
As it is well-known the presence of the first-class constraints in the theory
means the existence of a certain gauge symmetry.
Our analysis shows that in the light-cone Yang-Mills mechanics the original
$SU(2)$ gauge symmetry of the field theory, after supposition of the gauge
fields homogeneity, transforms into $SU(2) \times U(1) \times U(1)$ symmetry.


\section{
Computational aspects}
\label{sec:CA}



In the paper~\cite{GG99} a general algorithm for computing and separating constraints for
polynomial Lagrangians was devised.
The algorithm combines the constructive ideas of Dirac~\cite{DiracL} with the \Gr bases
techniques and called Dirac-\Gr algorithm.
Its implementation was done in Maple and in this section we characterize briefly the main
computational steps one needs to obtain the results of the previous section as they
were done by the Maple code.
In so doing the below described computational steps
is nothing else than concretization of the Dirac-\Gr algorithm to our model described in Sect.2.

Denote by $q_m$ and $\dot{q}_m$ $(1\leq m\leq 12)$, respectively, the generalized
Lagrangian coordinates in~(\ref{eq:lagr}) listed as
\begin{equation}
A^1_+\,, A^2_+\,, A^3_+\,, A^1_1\,, A^2_1\,, A^3_1\,, A^1_2\,, A^2_2\,, A^3_2\,, A^1_-\,, A^2_-\,, A^3_-
\end{equation}
and their velocities (time derivatives).
Then momenta are
\begin{equation}\label{eq:momenta}
p_m = \frac{\partial L}{\partial \dot{q}_m}\,, \qquad 1\leq m\leq 12 \,.
\end{equation}
To compute the primary constraints it suffices to eliminate the velocities $\dot{q}_m$
from the system (\ref{eq:momenta}) polynomial in $\dot{q}_m,\, q_m,\, p_m$.
The elimination are performed by computing a \Gr basis~\cite{CLO,BW93} for the generating
polynomial set
\begin{equation}
\{\ p_m - \frac{\partial L}{\partial \dot{q}_m}\ \mid \ 1\leq m\leq 12\ \}
\end{equation}
for an ordering (in Maple{\tt lexdeg}) eliminating velocities $\dot{q}_m$.
In the obtained set all algebraically dependent constraints~\cite{CLO} are ruled out.
Thus~(\ref{eq:prcon-1})-(\ref{eq:prcon-2}) is the algebraically independent set.

The canonical Hamiltonian~(\ref{eq:khrlh}) is determined as reduction of
\begin{equation}
p_m \dot{q}_m - L
\end{equation}
modulo the \Gr basis computed.
Then the computation of the Poisson brackets between the Hamiltonian variables
(generalized coordinates and momenta) as well as the computation of the total
Hamiltonian~(\ref{eq:toth}) is straightforward.

The next step is construction of the secondary constraints~(\ref{eq:secgauss})-(\ref{eq:secchi}).
It is done by reduction of the Poisson brackets of the primary constraints with
the total Hamiltonian modulo the set of primary constraints.
Again the \Gr basis technique provides the right algorithmic tool for doing such computations.
Thus, the complete set of algebraically independent constraints consists of twelve elements
\begin{equation}
\label{eq:complete}
\mathcal{F} =
\{
\varphi^{(1)}_a,\, \psi_k,\, \varphi^{(2)}_a, \chi^b_{k \bot},
\quad a, b = 1,2,3, \quad k=1,2\} \,,
\end{equation}
where from the six constraints $\chi^b_{k \bot}$ only four algebraically
independent are included in (\ref{eq:complete}) in accordance with the two
relations (\ref{eq:dependence}).

Next, to separate the complete set of constraints into first and
second classes one computes the $12\times 12$ Poisson bracket matrix on the constraint surface
\begin{equation}
M:=\|\, \{f_m\,,f_n\}_{\,\vert{CS}} \,\| \,,
\end{equation}
where $f_m,\, f_n \in \mathcal{F}$.
Since $\mbox{rank}\| M \| = 4$ the complete constraint set
$\mathcal{F}$
can be separated in four second-class constraints and eight first-class ones.
To select the first-class constraints it suffices to compute a basis
\begin{equation}
\mathcal{A} = \{\mathbf{a}_1,\ldots,\mathbf{a}_8\}
\end{equation}
of the null space for the matrix $\| M \|$ and then construct the first-class constraints as
\begin{equation}
(\mathbf{a}_s)_m f_m \,, \qquad 1\leq s \leq 8 \,.
\end{equation}
To extract the second-class constraints from $\mathcal{F}$ one constructs $8\times 12$ matrix
$\|\, (\mathbf{a}_s)_m\, \|$ from the components of the vectors in $\mathcal{A}$
and finds a basis
\begin{equation}
\mathcal{B} = \{\mathbf{b}_1,\ldots,\mathbf{b}_4\}
\end{equation}
of the null space of the constructed matrix.
Then every vector $\mathbf{b} \in \mathcal{B}$ yields a second-class constraint:
\begin{equation}
(\mathbf{b}_l)_m f_m\,, \qquad 1\leq l\leq 4\,.
\end{equation}
As a result, the eight first-class constraints are
$\left(\varphi^{(1)}_a, \psi_k, \varphi^{(2)}_a \right)$,
whereas four algebraically independent constraints from $\chi^a_{k \bot}$
are of the second-class.

Relations (\ref{eq:group_1})-(\ref{eq:group_4}), revealing the structure of the gauge group
generated by the first class constraints, can also be computed fully algorithmically.
To do this we extended of Maple package~\cite{GG99} with a general procedure that computes
the Poisson bracket of any two first-class constraints $\phi_r$ and $\phi_s$
as linear combination of elements in the set of first-class constraints:
\begin{equation} \label{eq:gauge_group}
\{\phi_r\,,\phi_s\} = c^q_{\ r s}\, \phi_q \,.
\end{equation}
With that end in view and in order to cope the most general case we implemented the extended
\Gr basis algorithm~\cite{BW93}.
Given a set of polynomials $Q = \{q_1,\ldots,q_m\}$
generating the polynomial ideal $<Q>$, this algorithm outputs the explicit representation
\begin{equation} \label{eq:gb}
g_n = h_{n m}\,q_m
\end{equation}
of elements in a \Gr basis $G = \{g_1\ldots,g_n\}$ of this ideal in terms of the polynomials
in $Q$.
Having computed a \Gr basis $G$ for the ideal generated by the first-class constraints and
the corresponding polynomial coefficients $h_{n m}$ for the elements in $G$
as given in~(\ref{eq:gb}),
the local group coefficients $c^q_{\ r s}$
(which may depend on the generalized coordinates and momenta) in~(\ref{eq:gauge_group})
are easily computed by reduction~\cite{CLO,BW93} of the Poisson brackets
modulo \Gr basis expressed in terms of the first-class constraints.

However, the use of this universal approach may be very expensive from the computational point
of view.
For this reason our Maple package tries first to apply the multivariate polynomial division
algorithm~\cite{CLO} modulo the set of first-class constraints.
Due to the special structure of the primary first-class constraints that usually include
those linear in momenta as in~(\ref{eq:prcon-1}), this algorithm often produces the right
representation~(\ref{eq:gb}); but unlike the extended \Gr basis algorithm does it very fast.
Correctness of the output is easily verified by computing of the reminder.
If the latter vanishes, then the output of the division algorithm is correct.
Otherwise the extended \Gr basis algorithm is applied.

In our case the division algorithm just produces the correct formulas
(\ref{eq:group_1})-(\ref{eq:group_4})
for the Poisson brackets of the first-class constraints
$\left(\varphi^{(1)}_a, \psi_k, \varphi^{(2)}_a \right)$.
Similarly, one obtains the formulas
(\ref{eq:bracket-1})-(\ref{eq:bracket-3}).


\section*{Acknowledgements}
\label{sec:Acnow}

It is a pleasure to express thanks to M.D. Mateev and V.P. Pavlov
for valuable discussions and suggestions.
The work was supported in part by the RFBR grant 01-01-00708.
Contribution of V.P.G. was also partially supported by the
RFBR grant 00-15-96691 and by grant INTAS 99-1222.
A.M.K. acknowledges INTAS for providing financial support, grant 00-00561.


\end{document}